# Nanoscale mechanical manipulation of ultrathin SiN membranes enabling infrared near-field microscopy of liquid-immersed samples

Enrico Baù[1], Thorsten Gölz[1], Martin Benoit[2,3], Andreas Tittl[1,*], and Fritz Keilmann[1]

1. Chair in Hybrid Nanosystems, Nano-Institute Munich, Faculty of Physics, Ludwig-Maximilians-University, Königinstr. 10, 80539, München, Germany

2. Chair of Applied Physics, Molecular physics of life, Faculty of Physics, Ludwig-Maximilians-University, Am Klopferspitz 18, 82152 Martinsried, Germany

3. Center for NanoScience, Ludwig-Maximilians-University, Amalienstr. 54, D-80799 München, Germany

**Abstract**

Scattering scanning near-field optical microscopy (s-SNOM) is a powerful technique for mid-infrared spectroscopy at nanometer length scales. By investigating objects in aqueous environments through ultrathin membranes, s-SNOM has recently been extended towards label-free nanoscopy of the dynamics of living cells and nanoparticles, assessing both the optical and the mechanical interactions between the tip, the membrane and the liquid suspension underneath. Here, we report that the tapping AFM tip induces a reversible nanometric deformation of the membrane manifested as either an indentation or protrusion. This mechanism depends on the driving force of the tapping cantilever, which we exploit to minimize topographical deformations of the membrane to improve optical measurements. Furthermore, we show that the tapping phase, or phase delay between driving signal and tip oscillation, is a highly sensitive observable for quantifying the mechanics of adhering objects, exhibiting highest contrast for low tapping amplitudes where the membrane remains nearly flat. We correlate mechanical responses with simultaneously recorded spectroscopy data to reveal the thickness of nanometric water pockets between membrane and adhering objects. Besides a general applicability of depth profiling, our technique holds great promise for studying mechano-active biopolymers and living cells, biomaterials that exhibit complex behaviors when under a mechanical load.

# 1. Introduction

Atomic force microscopy (AFM)-based [1] scattering scanning near-field optical microscopy (s-SNOM) [2, 3] using mid-infrared (MIR) light has become a promising tool for *in-situ* label-free imaging and spectroscopic studies of biological materials in water at resolutions far below the diffraction limit [4, 5]. Nanoscale microscopy is achieved by focusing MIR light onto an AFM probing tip **(Fig. 1a)**, whereby the tip shaft acts as an optical antenna that creates a self-focused and strongly confined hot spot below its apex. The achievable resolution of 20 nm [6] makes s-SNOM an excellent tool to image and spectroscopically characterize, for example, subcellular structures [7].

Importantly, s-SNOM operation using a broadband light source [3, 8] allows to record nanoscale complex-valued Fourier-transform infrared spectra (nano-FTIR) of a wide range of samples, including liquids [4, 5, 9], protein aggregates [10, 11] and dry [12] as well as in-liquid biological material [5, 13]. In general, s-SNOM can conveniently be used to identify any chemical compounds within a submicrometric spot of interest by taking local nano-FTIR spectra. The spectra are obtained in form of amplitude $s_n$ and phase $\varphi_n$ spectra acquired simultaneously, free of any background after applying demodulation at a suitably high harmonic (n>1) of the tip's tapping frequency $\Omega$ [14, 15]. The depth of near-field sensing into a material scales with the tip radius $r$. It is in the range of 100 nm for tips with r = 60 nm [5, 16], which allows near-field probing to reach through, for example, biological membranes into the interior of a cell.

Because s-SNOM is based on an AFM platform, it acquires simultaneously with the optical responses (**Fig. 1c**) the sample's mechanical responses (**Fig. 1b**), which are the surface topography and the mechanical phase delay $\varphi_{mech}$ between the signal driving the tapping and the actual oscillation due to the tip interacting with the sample. The latter can yield information on the sample's intrinsic elasticity, in addition to extrinsic forces due to charges or adhesion, as well as the energy dissipated by tapping [17]. Numerous AFM techniques have been used to study the mechanical properties of living cells in detail, such as cell-cell adhesion forces [18, 19], the mechanics of the cytoskeleton [20], density of receptors on a cell [21] and binding energies between ligands and receptors [22].

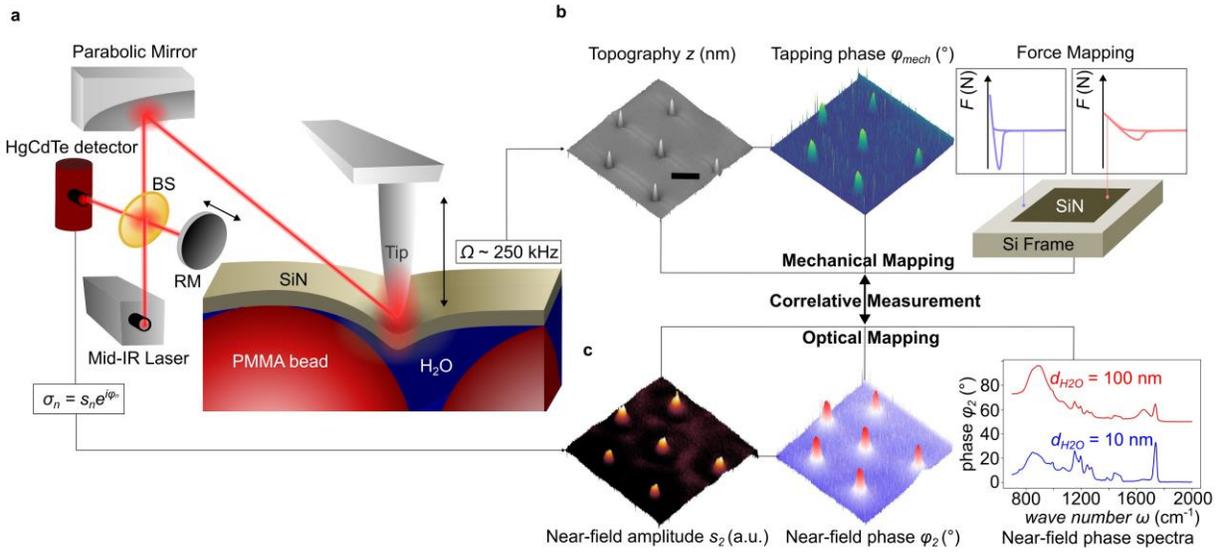

**Figure 1. Liquid s-SNOM setup and associated observables of interest. a** Schematic of SiN-membrane-based near-field probing into liquids using a metal tip operated in tapping mode, able to map the samples's MIR optical response, simultaneously with the mechanical deformation of the membrane caused the probing tip and objects adhering to the membrane. **b** The mechanical channel senses the deflected laser beam (not shown) and returns topography $z$ and phase delay $\varphi_{mech}$ between cantilever drive and tapping motion. Additionally, force-distance curves can be taken on selected points to quantitatively determine a material's elastic properties. **c** The optical channel senses back-scattering of the illuminating laser via a Michelson interferometer, which contains a beamsplitter BS and a movable reference mirror RM, and returns an optical amplitude and phase (for instrumental details see **Experimental Section**, for observed results in **b**, **c**, see main text).

For s-SNOM of samples in water (**Fig. 1a**), the use of a SiN membrane has three important, unique advantages, mainly that (i) it is highly transparent to the probing MIR near field (depicted as a transparent red ellipsoid), (ii) it protects samples from drying by water evaporation (or from contamination/infection through atmospheric constituents), and (iii) it provides a stable, flat and homogenous surface for the scanning tip [5]. As demonstrated in our previous study on *in-situ* nano-imaging of living cells and microparticles adhering in water below a 10-nm thin SiN membrane, such objects can induce additional, topographical contrasts as well as tapping phase contrasts through nanoscale deformations of the membrane [5]. Since our s-SNOM (see **Experimental Section**) stores all optical and mechanical observables simultaneously, it serves as a highly correlative nano-imaging technique.

Here we assess the mechanics of the flexible, 10 nm thin membrane freely spanned across a solid frame in contact with water by measuring the reversible indentation of an AFM tip through force-distance curves, both in contact and tapping mode, the latter with a systematic variation of tapping parameters such as the tapping amplitude. We further investigate the case where submerged, spatially fixed and well-defined PMMA spheres with 10 μm diameter adhere and support the membrane by recording mechanical and optical images. Despite the SiN membrane fully covering the sample, we find that a wealth of information can be extracted by considering the deformation of this thin covering sheet.

Furthermore, apart from extracting the system's effective elasticity, we show theoretically and experimentally that the SiN membrane deformation is controllable through a simple variation of the tapping parameters. We find that the tapping phase $\varphi_{mech}$ is the critical observable to determine the mechanical properties of a sample adhering to the thin SiN membrane and show that lower tapping amplitudes are preferable to maximize tapping phase contrast while minimizing topographical deformation.

Lastly, we correlate mechanical and optical data to enable depth profiling of thin water pockets that form in between sample and membrane, a technique which could in the future be straightforwardly extended to other, more complex biological materials, such as proteins or lipids, which could be studied in more detail in order to investigate their prompt response to external manipulation, for example, by changing the mechanical pressure exerted by the AFM tip, or by changing the pH value or the salt concentration of the solution.

## 2. Results and Discussion

### 2.1. Mechanical interactions between s-SNOM tip and SiN membrane.

To understand the mechanics of the SiN membrane, we analyze the indentation caused by the tapping tip. In general, the motion of an AFM tip interacting with a surface can be described as a driven damped oscillator system, taking into account attractive and repulsive forces between tip and sample [23]. These forces include long-range Van-der-Waals, electrostatic and short-range repulsive forces and are heavily dependent on the sample and tip material properties.

We recorded force-distance curves in contact mode on a 10 nm SiN membrane either with liquid underneath or on the rigid Si-frame (**Fig. 2**). The deproach curves were taken for a duration of 60s each (velocity of 17 nm/s), with the initially applied force set to 800 nN on Si, and to 500 nN on water. A sketch of the experimental system is shown in **Fig. S1**. On liquid, the SiN membrane is initially deformed to approximately 150 nm depth (brown curve in **Fig. 2**), demonstrating that the tip reversibly induces an approximately 150 nm deep dimple in the water-supported suspended

SiN membrane. In the case of a free-standing membrane without any water-support, the shape of such dimples has been shown to depend strongly on the membrane's in-plane tension [24]. Our curve taken on Si, on the other hand, merely reflects the cantilever's bending, which can be used to correct the curve taken on water and to determine the correct indentation depth. The pull-off force on water measures 25 ± 3 nN.

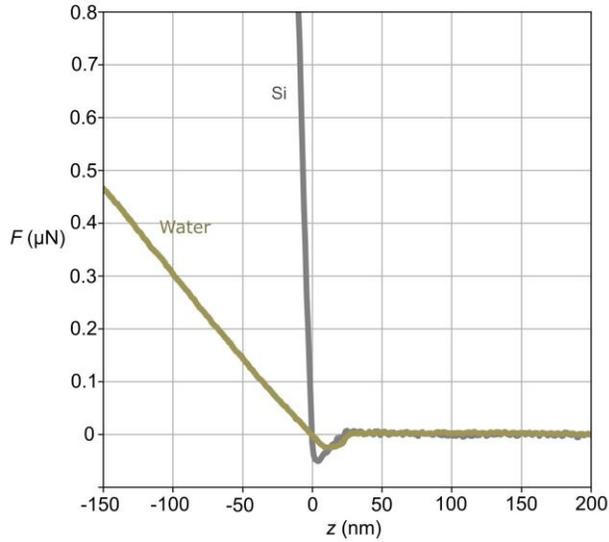

**Figure 2. Contact-mode deproach from a 10-nm thick SiN membrane** pre-stretched at <250 MPa over an opening of 250x250 µm² with water adhering from below. Experimental deproach curves in contact-mode AFM taken sequentially on the Si frame (gray) and on water (20 µm away from the frame, brown), during 1 min each, the latter demonstrating, at the start, an approximately 150 nm deep, tip-induced dimple in the membrane.

If we consider the sample simplified as an elastic, homogenous half-space, its effective Young's modulus could be computed from a parabolic fit to the measured force-distance curves using the Johnson Kendall-Roberts (JKR) model [25, 26], assuming a pyramid-shaped indenter. Applying this model to the SiN/water system, we find that $E_{eff}$=30±10 MPa, a value that is four orders of magnitude smaller than measured on the membrane-covered Si-frame. Force-volume measurements (**Fig. S2**) yield a similar value between 20-50 MPa on the SiN/water region. In addition, they show that the effective stiffness decreases away from the Si border frame. Predicting the exact shape of this deformation would necessitate numerical simulations (as has been done previously in for stretched membranes in vacuum [24]), which would have to be extended to account for the viscoelastic properties of the confined water.

For tapping-mode deproach curves, the initial in-contact tapping amplitude *a* is the decisive parameter. In most s-SNOM experiments, it is chosen between 20 and 100 nm. Typical tapping frequencies are 250-350 kHz. The effective force applied to the membrane by the oscillating tip is an average over changing forces encountered during an oscillation cycle. By contrast, when taking contact-mode force-distance curves, the tip interacts with the surface for much longer durations, thus allowing short-range repulsive forces to have greater impact.

Tapping-mode deproach curves were taken on the same sample as contact-mode curves shown previously, either on Si or on water at 90% of the free cantilever's resonance frequency $\Omega_0$ and at a set point of 85% of the free cantilever's amplitude $a_0$. For a given tapping amplitude *a*, the measurements of all four observables were taken simultaneously (**Fig. 3**).

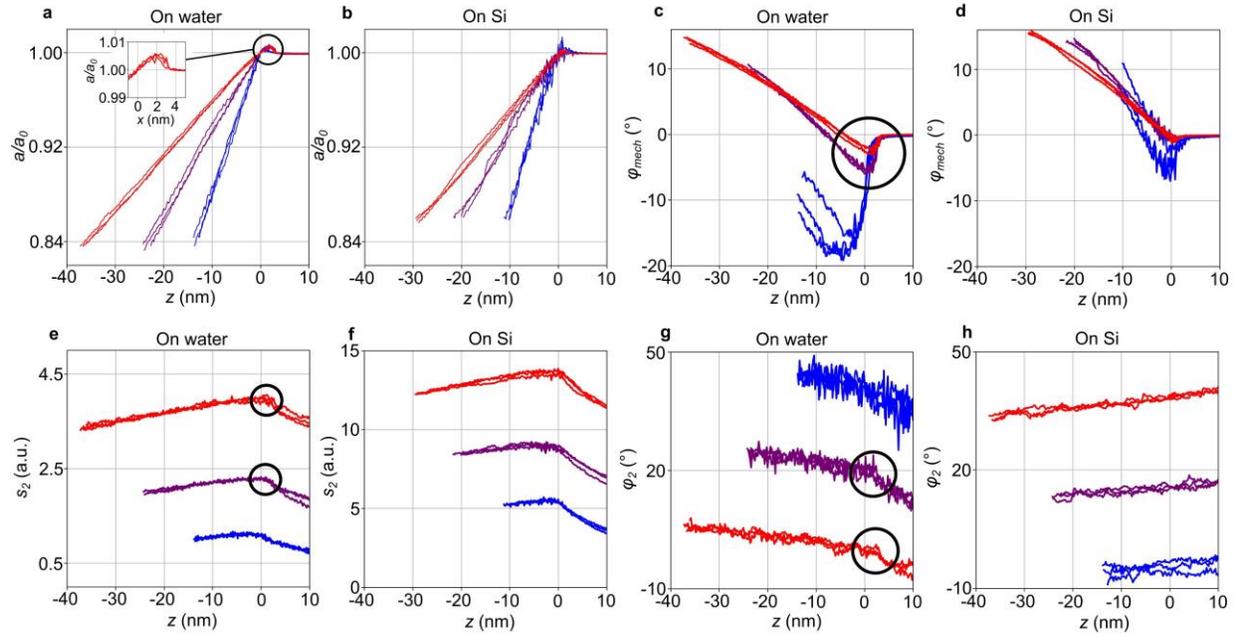

**Figure 3. Tapping-mode deproach curves from the 10-nm SiN membrane either on water or on Si frame. a, b** Tapping amplitude $a/a_0$, **c, d** tapping phase $\varphi_{mech}$, **e, f** optical amplitude $s_2$ (at the peak of water absorption, 1650 cm$^{-1}$) and simultaneously recorded **g, h** optical phase $\varphi_2$ at three different tapping amplitudes $a = 60$ nm (blue), $a = 120$ nm (purple) and $a = 200$ nm (red). The zero on the z-axis is set to where the amplitude *a* reaches $a_0$, i.e. its constant value beyond the loss of contact. Anomalies (encircled, inset) suggest a repeatable tip-induced bulging upwards of the membrane.

The curves taken on water (**Fig. 3a**) show an unexpected effect, which is clearly not present in those taken on the Si frame (**Fig. 3b**): the tapping amplitude *a* rises above $a_0$ for $x > 0$, indicated by a black circle and an inset). Our interpretation is that this deproach region is dominated by adhesion forces from the tip that pull the membrane upwards until it snaps back. The same effect becomes visible also in the simultaneously recorded tapping phase $\varphi_{mech}$ (encircled in **Fig. 3c**), which shows an even more abrupt snap back. This observation adds to our former demonstration with living cells [5] that the tapping phase is especially sensitive for mapping local contrasts of adhering objects. In addition, the optical amplitude (**Fig. 3e**) and optical phase (**Fig. 3g**) keep developing continuously with $a$ past $x = 0$, until they experience abrupt changes in behavior (black circles), as can be expected for a sudden opening of an air gap between tip and sample. Measurements on the Si frame (**Fig. 3b, d, f, h**) do not exhibit the same trend.

**2.2 S-SNOM mechanical nano-imaging of submerged objects**

Our former experiments have revealed that a thin membrane covering water deforms under the mechanical load of an AFM tip [5], an effect demonstrated across the edge of the Si-frame at a single tapping amplitude $a$. Here we study how this deformation depends on $a$ using an aqueous sample with the membrane supported by PMMA spheres (10 or 2 μm diameter) close to the Si frame's edge (**Fig. 4**). With $a/a_0$ set to 0.8, we acquire a series of AFM images (topography $z$ and tapping phase $\varphi_{mech}$) for $a$ between 40 and 200 nm (**Figs. 4a-f**). The PMMA spheres are arranged in hexagonal patterns, each creating a bulge in the membrane surface, as can be seen in the topography images. We conclude that the spheres form self-organized 3-d hexagonal crystals fixed to the Si frame and to the supporting the pre-tensioned membrane. The crystals seem strong enough that the tip's tapping forces do not affect their structure.

The tapping forces clearly induce a reversible dimple in the membrane at almost any point of probing. The topography in **Figs. 4a,c** appears flat between spheres (of both sizes), from which we deduce that the dimple's depth is constant on deep enough water. However, at the points where the spheres are in contact with the membrane, the lattice structure provides stability and thus reduces the membrane's flexibility. At the centre of each sphere, the deformation fully vanishes, which can be concluded from the fact that the topographical height $z$ there becomes independent of the tapping amplitude $a$. Altogether, the PMMA lattice supports the pre-tensioned membrane and enables to extract information on tip-induced topographical deformations.

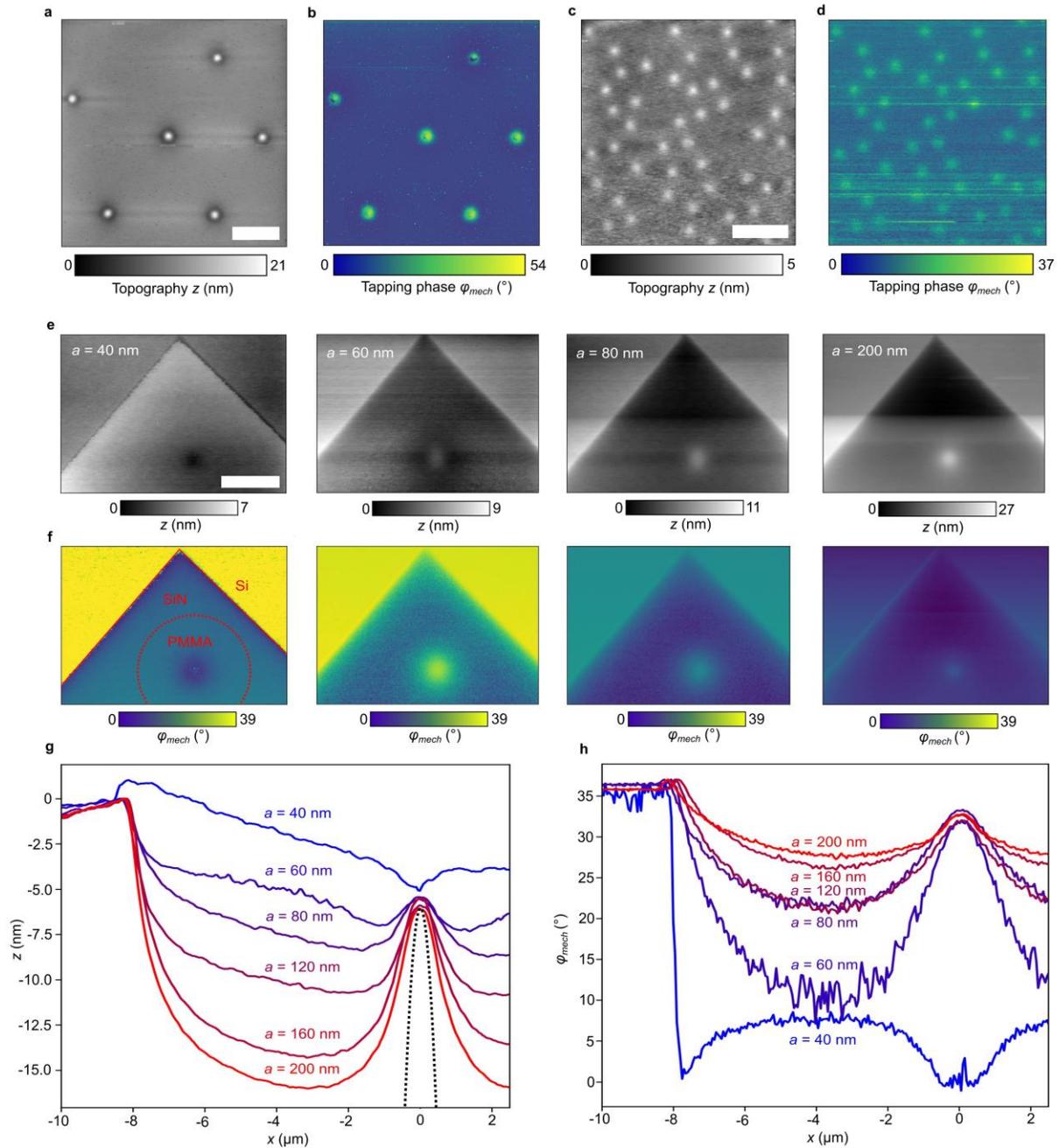

**Figure 4. Mechanical maps of hexagonally arranged PMMA microspheres in water a, c** Topography $z$ and **b, d** tapping phase $\varphi_{mech}$ for spheres of diameter 10 µm and 2 µm respectively, adhering below a 10 nm thin SiN membrane. **e** Topography and **f** tapping phase images of a single 10-µm PMMA sphere near a Si corner, from left to right at $a$ = 40 nm, 60 nm, 80 nm, 200 nm. All scale bars are 5 µm. **g** Topography profiles and **h** mechanical profiles of $\varphi_{mech}$ extracted from e,f demonstrate that the tip is pulling the membrane upwards when $a$ is chosen smaller than 60 nm. The dotted curve in g depicts the shape of a 10 µm PMMA sphere.

The mechanical tapping phase $\varphi_{mech}$, on the other hand, effectively measures the total dissipated energy at each pixel of a sample's surface [27] (see **SI Section 3** for details). **Figs. 4b, d** and **f** show that $\varphi_{mech}$ increases on stiffer samples, in agreement with our earlier report (Fig. 2c in [5]). Remarkably, decreasing tapping amplitudes results in increasing the $\varphi_{mech}$ contrast between an adhering particle and its aqueous surrounding (**Fig. 4h**). This observation clearly shows that low tapping amplitudes enhance the mechanical sensitivity of detecting local changes of elasticity. **Fig. 4g** further shows that topographical deformations can be almost fully eliminated just by choosing a suitably small tapping amplitude. The achievement of a structureless topography would, in principle, suppress any topography-induced crosstalk in optical near-field images. In addition, the use of small tapping amplitudes is a well-known advantage of better suppressing optical background-scattering signals [2].

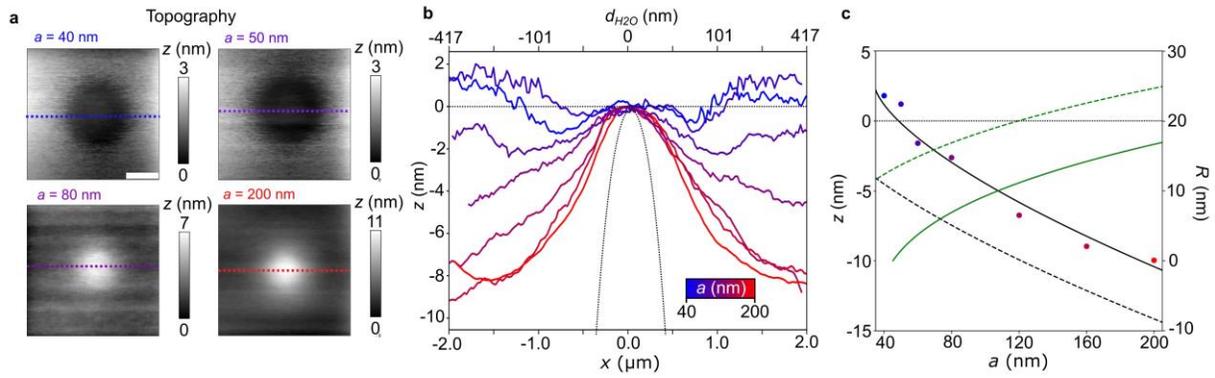

**Figure 5. Topography reversal at low tapping amplitudes. a** Topography of a single 10 μm PMMA sphere in water at different values of *a* as indicated, Scale bar 1 μm. **b** Profiles extracted from dotted lines vs. *x* (lower axis), or respectively, nominal water thickness $d_{H2O}$ (upper axis), defined by $d_{H2O} = R_{PMMA} - \sqrt{R_{PMMA}^2 - x^2}$, where $R_{PMMA}$ is the sphere radius. The dotted curve depicts the PMMA sphere. **c** Extracted height difference *z* between the sphere center and far away (2 μm), for *a* between 40 nm and 200 nm. Full and dashed curves, respectively, show the calculated fitted indent using the Hertz (dashed) and JKR (solid) model. Green curves show the respective indent radius *R* calculated with both models (see **SI Section 4** for more information).

**Fig. 5** shows the tip-induced membrane deformation in more detail for a single 10 μm PMMA sphere. The membrane exhibits a dimple or a bulge deformation depending on *a*. For *a* = 60 nm, the membrane stays approximately flat, while for lower tapping amplitudes, the membrane is being pulled upwards. In **Fig. 5c**, we extracted bulge/dimple heights *z* (relative to surrounding water) vs. *a*. The solid curves show the results of analytical JKR model calculations, where the adhesion energy was fitted to match our experimental data. The contact radius *R* was calculated using equation (3) (as described in **SI Section** 4). The dashed curve shows corresponding results obtained with the analytical Hertz model, which neglects attractive forces (see **SI Section 4** for details).

It is important to note that the SiN membrane's upper surface may be covered by a thin water film, which could, through capillary forces, strongly attract the tip to the surface [28]. This mechanism may explain why we are not able to achieve stable tapping with *a* below 20-30 nm, as the tip gets trapped on the surface and is unable to oscillate properly, or even crashes into the membrane. The mechanism could also be responsible for the observed increased fluctuations in the measured deproach curves at small tapping amplitude (**Figs 3b** and **3d**, blue for a=60 nm).

The full set of images from which the data in **Fig. 5b** has been extracted is shown in **Fig. S3**. Surprisingly, the topographies at *a* = 40 and 50 nm exhibit rings which display a shallow minimum around the sphere's centre. These rings may indicate the highly interesting existence of a critical thickness of confined water on the orders of tens of nm, below which the attraction of the membrane by the tip is overcompensated through adhesion forces between membrane and PMMA sphere. While the physical origin of this observation is beyond the scope of the present study, we tentatively suggest that (i) a rheological contribution could become important with ultrathin water layers, and that (ii) the widely observed changes of the water's supramolecular structure next to surfaces may play a role.

The effect of increasing membrane deformation for higher tapping amplitudes almost fully vanishes when the sample is dried out (**Fig. S4**). Then, the particle forms a permanent bulge of around 80 nm height, irrespective of the tapping amplitude used for imaging. We attribute this bulge to non-volatile contaminants of the PMMA suspension which accumulate during the process of drying on the underside of the membrane, thereby permanently deforming the membrane. This also explains the strong local deformations of the membrane visible across the entire topographical scan in **Fig. S4a**, as opposed to the relatively flat surface seen in prior measurements.

In order to find the tapping parameters yielding maximum phase contrast, we varied both tapping amplitude *a* and amplitude setpoint $a/a_0$ in a 3x3 matrix of mechanical images of the same PMMA sphere (**Fig. 6**). It is well known in the AFM literature that decreasing the amplitude setpoint increases the tip's time-averaged loading force [29] and results in detuning the driving frequency $\Omega$ further away from the cantilever resonance frequency $\Omega_0$. Additionally, as can be seen from quantitative, extracted profiles shown in **Fig. S5**, lower amplitude setpoints induce deeper dimples in the membrane away from the sphere, as do higher tapping amplitudes *a*. On the other hand, higher setpoints and lower tapping amplitudes both lead to higher tapping phase contrasts while keeping the membrane almost flat.

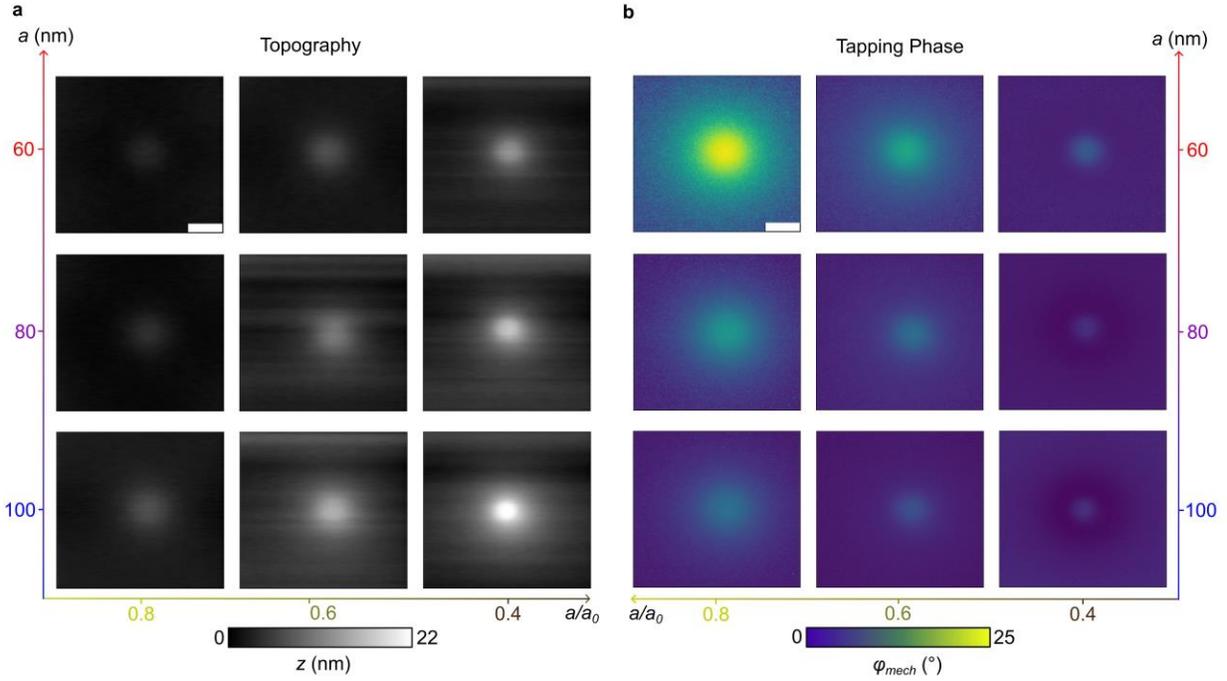

**Figure 6. Influences of tapping amplitude and setpoint. a** Measured topography and **b** tapping phase map of a 10-µm PMMA sphere in water adhering underneath a 10-nm SiN membrane, assembled in a 3x3 matrix with varied tapping amplitudes (vertical) and amplitude setpoints (horizontal). Extracted profiles are shown in **Fig. S5**.

### 2.3. Correlating Infrared Spectroscopic and Mechanical Information

Finally, we extend our previous approach of distinguishing stacked sub-membrane layers [5] by recording numerous nano-FTIR phase spectra $\varphi_2$ along a radial line, starting from a 10-µm PMMA sphere's center, for two different tapping amplitudes (**Figs. 7a**, **b**). Quantitative complex spectra on PMMA and on water shown in **Fig. S6** were used to define a characteristic frequency, 1735 cm$^{-1}$ and 1650 cm$^{-1}$, for each material respectively. We extracted phase signal heights at both characteristic frequencies from each spectrum in **Figs. 7a**, **b** and plotted the spectral weight of both materials vs radial position $x$ shown in **Figs. 7c**, **d.**

For a theoretical prediction, we first assume a plane membrane touching the sphere at a single point, from which it follows that the nominal water thickness $d_{H2O}$ increases with x, as already given in the caption of **Fig. 5**. The finite dipole model of s-SNOM [30, 31, 32] extended to a 3-layer sample (see **Experimental Section**) then yields the dependencies shown in **Figs. 7c**, **d** (full curves, phases at $x$=0 were multiplied by 1.05 to match the measured values).

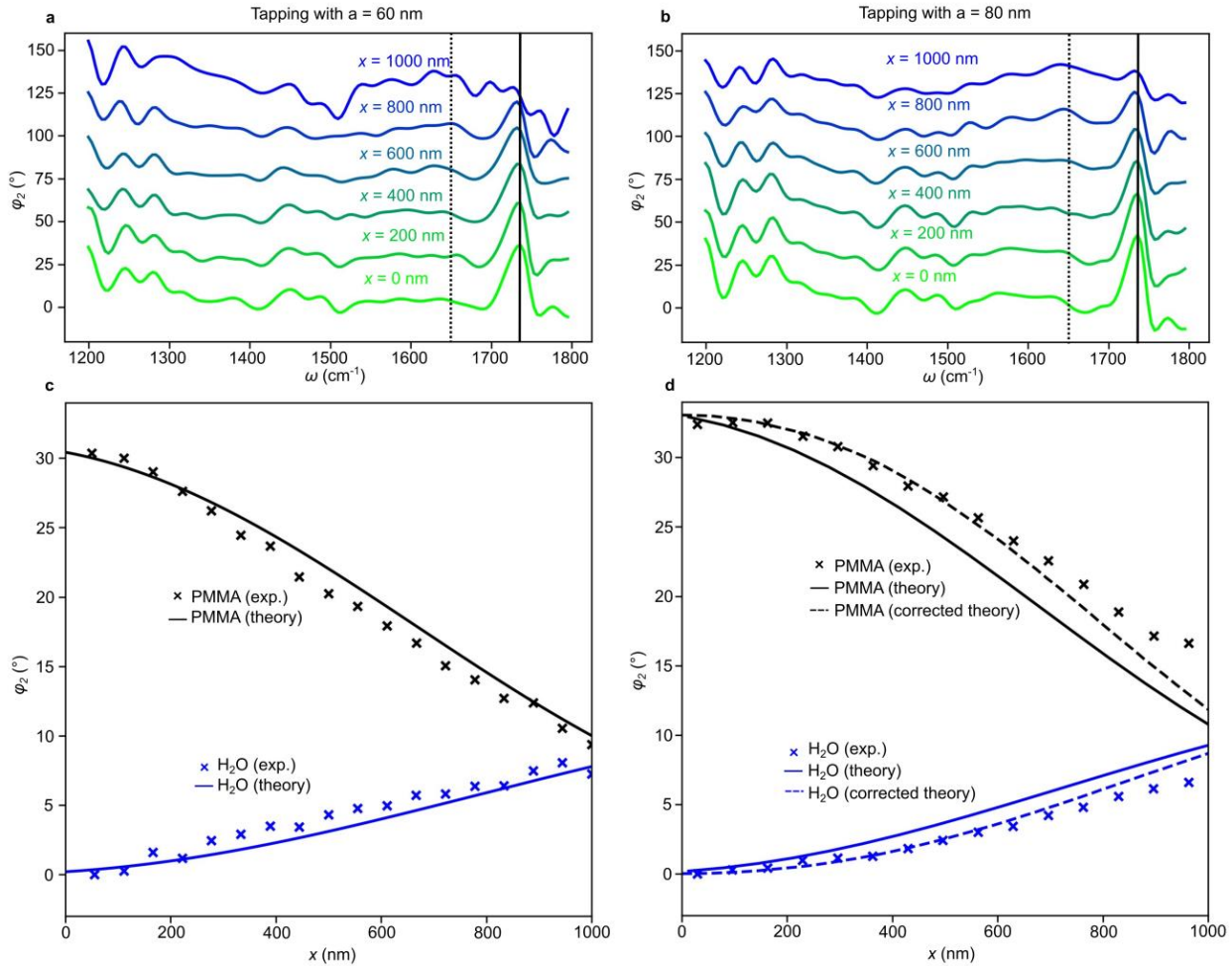

**Figure 7. Topography-corrected subsurface near-field probing. a, b** $\varphi_2$ spectra taken at radial distances x from the starting position ($x$=0) above a 10-µm PMMA sphere's center, using $a$ = 60 nm or 80 nm respectively, show key vibrational resonances assigned to carbonyl (at 1735 cm$^{-1}$, solid line) and to the bending vibration of $H_2O$ (at 1650 cm$^{-1}$, dotted line). Curves are offset for clarity by 25° each. **c, d** Plots of extracted peak $\varphi_2$ signals for PMMA (black crosses) and water (blue crosses) vs. radial distance $x$, using $a$ = 60 nm or 80 nm respectively, and calculated theory curves (full and dashed curves, see text for details).

Interestingly, the theoretically calculated curves match the measured points better at $a$ = 60 nm than at $a$ = 80 nm. For an explanation, recall from **Figs. 4, 5** above that for $a$ = 60 nm, the membrane is almost completely flat, while for $a$ = 80 nm, a sizable dimple forms outside the sphere center, which means that the thickness of the water space there is smaller than $d_{H2O}$. Therefore, the signal of the carbonyl resonance at 1735 cm$^{-1}$ comes out higher than theoretically expected. Since the deformation of the membrane is known from recorded topography images, we can calculate a corrected theoretical prediction by simply replacing $d_{H2O}$ by $d_{H2O} - z$, and indeed obtain curves that fit the data much better (dashed in **Fig. 7d**). Clearly, using topographical data to analyze infrared-spectroscopic measurements of multilayer samples can be of great value and should, in particular, enable accurate depth profiling of processes in aqueous environments.

Correlating infrared-spectroscopic nano-imaging with AFM data not only works for rigid PMMA spheres, but can certainly be applied to more elastic biomaterials. As an example, we present measurements on aggregated amyloid-beta protein strands in water in **Fig. S7**. Here the tapping phase exhibits a lower value on the object than in its surroundings, probably due to the softness of the protein aggregate. Interestingly, the topography is repeatably measured to be almost fully flat over multiple hours. Additionally, the aggregated proteins are clearly visible in the near-field optical signal simultaneously recorded, illustrating the potential to use this technique for various kinds of biological and chemical applications in the future.

## 3. Conclusions

Our measurements of membrane topography and tapping phase nano-images, supported by analytical modeling, clearly show that the tip-induced deformation of a thin SiN membrane separating air and water half-spaces can be controlled by two tapping parameters, amplitude $a$ and setpoint $a/a_0$. Unexpected features seen in tapping-mode deproach curves reveal a novel mechanism of manipulating the membrane, pulling it upwards and forming a water-filled bulge. This mechanism is simultaneously documented in the development of four different mechanical and optical observables. Abrupt changes in these developments when the tip reaches a certain height indicate a sudden appearance of an air gap between tip and membrane due to the collapse of the bulge when the membrane looses contact to the tip and snaps back. Furthermore, our measurements reveal that the tapping phase $\varphi_{mech}$ is the most sensitive observable to detect mechanical changes and thus represents the key for unlocking new insights on the mechanical behavior of adhering samples in water, such as migrating living cells.

Furthermore, our AFM imaging of hexagonally arranged PMMA spheres in water provides a guide-line to achieve highest tapping phase $\varphi_{mech}$ contrast: small tapping amplitudes and high setpoints. A welcome side effect is that small tapping amplitudes and high setpoints minimize tip-induced deformations of the covering SiN membrane, which is highly desirable to avoid topographical artefacts in simultaneously recorded near-field optical signals. Accomplishing high tapping phase contrasts for in-liquid samples would allow for future studies of highly interesting dynamic processes, such as the mechanics of cell adhesion to elastic surfaces or local variations of stiffness on the surface of a living cell when injected with chemicals that alter their architecture [33, 34, 35].

Infrared near-field probing through the membrane and a thin water layer into a PMMA sphere adhering to the membrane allows to quantitatively assess spectra as well as the local thickness of the water. Using this technique, future experiments may contribute to our knowledge of structure and dynamics of asymptotically thin water spaces, enabling s-SNOM to be utilized as a tool for subsurface tomography, as has been done in previous studies with dry samples [36, 14]. We also observed that nano-FTIR is able to resolve and characterize single sub-micrometer protein aggregates in water (**Fig. S7**), which are much less rigid than PMMA spheres. This capability could potentially unlock the key to a long-standing open question in biology, of revealing the mechanism (including dynamics) of a prion infection at the single-molecule level [37, 38]. Notably, membrane-based liquid s-SNOM has the potential for enabling a new class of *in-situ* experiments, by straightforwardly adding a microfluidic system. Hereby, the liquid medium in the sample cell could be exchanged to vary, for example, its pH value, its salt concentration or its content of specific biochemical agents, in principle without disturbing the adhering objects' position, orientation, and nanometer-scale resolved morphology.

## 4. Experimental Section

**AFM-based scattering scanning near-field optical microscopy**

In this work we use a commercially available s-SNOM (NeaSNOM from *attocube systems,* Haar, Germany), operated in nano-FTIR mode. Hereby, the tip (nano-FTIR tip, from *attocube systems*) is illuminated by a broadband MIR coherent source based on difference-frequency generation of <100 fs fiber laser pulses (FemtoFiber dichro midIR NEA 31002 from Toptica Photonics). The source emits an 800 cm$^{-1}$ wide spectrum and is centered at 1300 cm$^{-1}$. The optical path of the s-SNOM features a ZnSe beamsplitter, a paraboloidal focusing mirror, a liquid nitrogen-cooled HgCdTe-detector (IRA-20-00103 from InfraRed Associates) and a second parabolic mirror for focusing the back-scattered light onto the detector's active area of 0.0025 mm$^2$. The beamsplitter (BS) sends part of the incident beam to a reference mirror (RM) on a moving stage to form an asymmetric Michelson interferometer, where it interferes with the back-scattered beam. The resulting interferogram yields a complex-valued scattering coefficient $\sigma_n$ containing both amplitude $s_n$ and phase $\varphi_n$ information, demodulated at the nth harmonic of the tip oscillation frequency $\Omega$ in order to eliminate background-scattering signals. This allows to either record the total scattered light from the tip (white light imaging), or to acquire local nano-FTIR spectra by periodically moving the RM.

**Sample preparation in liquid cell**

The liquid cell used in this work [5] consists of a metal plate with a central hole of 2 mm diameter that is covered by fixing, with double-sided adhesive tape, a commercial Si chip carrying a freestanding SiN membrane (NX5025Z, Norcada, Edmonton, Canada). The membrane is 250x250 µm² in size and 10 nm thick. The membrane is UV-irridated for 30-40 minutes, to ensure sufficient hydrophilicity of the SiN surface for the liquid sample to wet the surface of the membrane. Subsequently, around 20 µL of a 1:10 diluted 10 µm-diameter PMMA microsphere solution (PMMA-F-10.0, microparticles GmbH, Berlin, Germany) is dropcasted into the metal holder and incubated for 15 min. The adherence of microparticles and water to the membrane is checked by visual inspection with an optical microscope. In the end, the chip is sealed with a glass cover slide to prevent water from evaporating.

**Computations and data processing**

All analytical calculations and data processing of experimental results in this work have been computed using Python. The JKR model was used for all calculations of tip-induced mechanical, effective indentation depths and contact radii. To predict the optical response of multilayered materials analytically, the finite dipole model was employed, extended to multi-layer objects by the transfer matrix method [31]. It is an excellent and computationally effective tool that requires only knowledge of the tip geometry and the layer's dielectric function, in our case for SiN [39], water [40] and PMMA [41]. For our calculations, the PMMA particles are assumed to have perfect spherical geometry. The model parameters used are $L = 300$ nm, $g = 0.6$, $r = 60$ nm.

**Acknowledgments**

This work was funded by the Deutsche Forschungsgemeinschaft (DFG, German Research Foundation) under grant numbers EXC 2089/1 – 390776260 (Germany's Excellence Strategy) and TI 1063/1 (Emmy Noether Program) and the Center for NanoScience (CeNS). Funded by the European Union (ERC, METANEXT, 101078018 and EIC, NEHO, 101046329). Views and opinions expressed are however those of the author(s) only and do not necessarily reflect those of the European Union, the European Research Council Executive Agency, or the SMEs Executive Agency (EISMEA). Neither the European Union nor the granting authority can be held responsible for them.

**Author contributions**

E.B. prepared samples and performed s-SNOM and nano-FTIR experiments. F.K. and T.G. conceived the idea and implemented the SiN liquid cell. E.B., T.G. and F.K. analyzed the data. E.B. conducted model calculations. E.B., M.B., A.T. and F.K. interpreted the results. E.B. and F.K. wrote the paper. All authors contributed to the scientific discussion.

**Conflict of interest**

F. K. is a scientific advisor to attocube systems AG which manufactures the s-SNOM used in this study. T. G. obtained financial support for his PhD thesis from attocube systems.

**Supplementary Information**

## 1. Mechanics of an AFM tip interacting with an elastic half-space

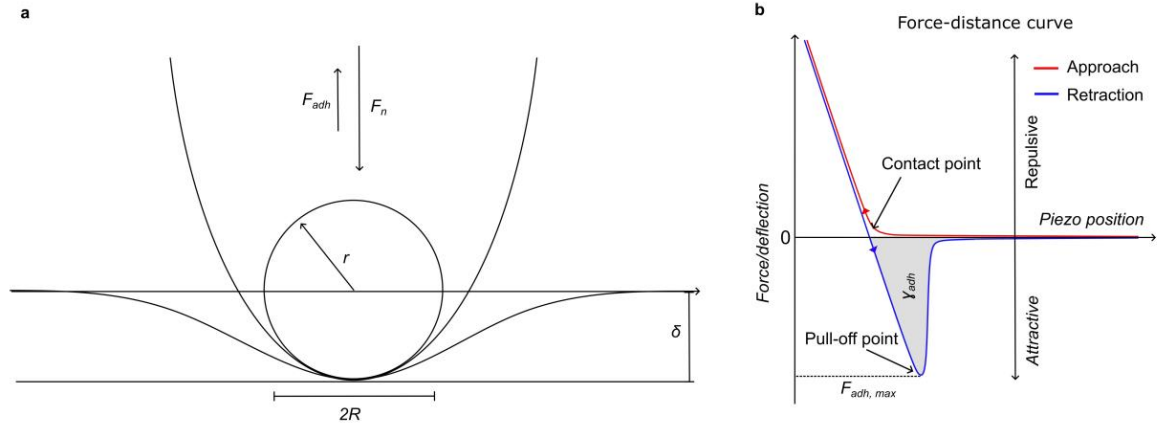

**Figure S1. Tip-sample interaction and force-distance curves a** Schematic of an AFM tip, approximated as a sphere with a radius of *r*, equivalent to the apex radius of the tip. The AFM tip is being propelled into the membrane with a normal loading force $F_n$. The AFM tip indents the SiN membrane to an indentation depth $\delta$. The radius of the resulting contact area is denoted *R*. Additionally, long-range adhesion forces between tip and sample surface occur, here denoted $F_{adh}$. **b** Schematic force–distance curve. During the approach (red) of the AFM tip towards the sample surface, the contact point will be reached, leading to a steady increase in repulsive forces. The retrace curve (blue) shows that, after passing a point where both attractive and repulsive forces balance each other, the tip enters an attractive regime, until reaching the pull-off point, where contact between tip and sample is lost. The grey area under the curves yields the adhesion energy and the maximum adhesion force is determined at the pull-off point.

The equation of motion of a tapping AFM probe can be expressed as:

$$m\frac{d^2z}{dt^2} + \frac{m\omega_0}{Q}\frac{dz}{dt} + kz = F_0 \cos \omega t + F_{TS} \qquad (1)$$

where *m* denotes the mass of the probe, $\omega_0$ its resonance frequency and $\omega$ frequency when in contact with the sample, *Q* the quality factor, *k* the spring constant of the oscillating AFM cantilever, $F_0$ the driving force and $F_{TS}$ the sum of interaction forces between tip and sample. A schematic of this theoretical system is shown in **Fig. S1a**. The AFM tip tapping from above deforms the sample, indenting it until it reaches its lowest point, as shown in **Fig. S1b** (red curve). Subsequently, the AFM tip is retracted (blue curve) until the indentation levels off at the $F = 0$ nN line. Beyond this point and until the pull-off point is reached, adhesion forces dominate. The greyed-out area in between the $F = 0$ nN line and retraction curve estimates the total work of adhesion. The maximum adhesion force is reached at the pull-off point.

## 2. Tip force measurements of wet membrane/Si interface

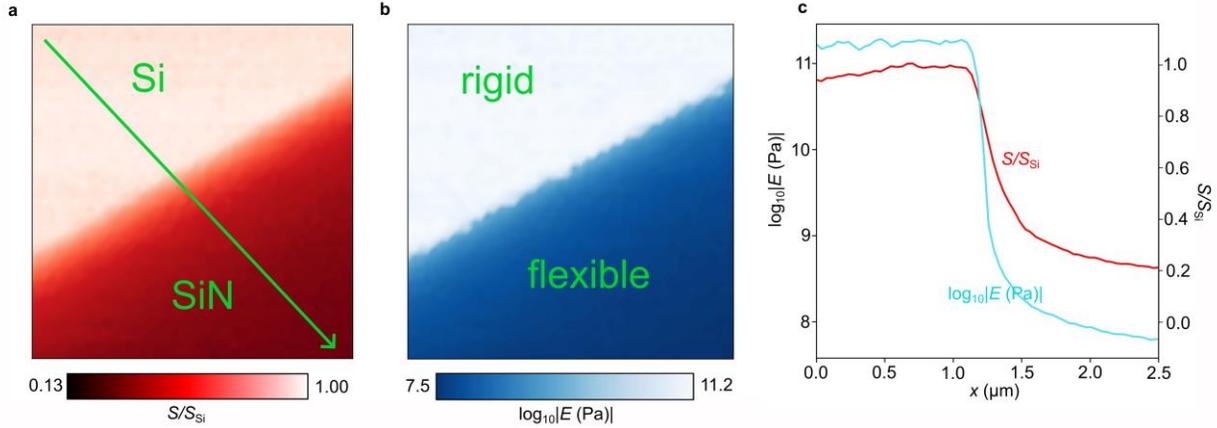

**Figure S2. Tip force measurements a** Stiffness referenced to Si and **b** Young's modulus (plotted logarithmically) of the Si/SiN membrane interface in the presence of water. The quantities were extracted by taking force-distance curves at each pixel and fitting the Hertz model to each curve. **c** Profiles of stiffness and Young's modulus (plotted logarithmically) across the SiN membrane/Si border extracted from a, b (green).

## 3. Tapping phase

Phase contrasts in AFM are registered by measuring the difference between the phase angle of the excitation signal and the phase angle of the cantilever response at each position. Phase shifts are associated with the presence of inelastic tip-sample interactions, which establishes a relation between the sine of the phase shift and the power loss. The external energy $E_{ext}$ steadily supplied to the cantilever equals the energy dissipated via hydrodynamic viscous interactions with the environment (air in our case) and via tip-sample interactions, $E_{ext} = E_{air} + E_{TS}$, where $E_{TS}$, is the energy dissipated through tip-sample interactions. From this, the mechanical phase $\varphi_{mech}$ can be described with the following equation [17]:

$$\sin \varphi_{mech} = \frac{\omega}{\omega_0} \frac{a}{a_0} + \frac{QE_{TS}}{\pi k a a_0} \tag{2}$$

## 4. Johnson Kendall-Roberts model

Calculating exact quantitative adhesion forces involved in tapping mode AFM requires extensive computations [42] Previous publications [24, 43, 44, 45] dealing with stretched thin sheets have treated the tip-sample interactions using either numerical computations or simplified analytical versions of Kirchhoff plate theory which takes into account the pre-stress applied to the membrane after fabrication. However, since we consider a layer of water below the SiN sheet, such thin-film models may be less applicable in our case. With a membrane of area of 250x250 μm², thickness of 10 nm, and a pretension of 250 MPa (as specified in the datasheet), the Kirchhoff plate model yields a mostly linear behavior of the indentation vs. applied force, which does not correctly describe our experimental findings shown in **Fig. 5**. Additionally, the model is not able to explain the membrane being pulled upwards by the tip for smaller tapping amplitudes, also seen in **Fig. 5**.

On the other hand, while the Hertz model of contact mechanics [46] is the simplest solution to solve most contact mechanics problems, it doesn't take into account the attractive forces between tip and membrane, thus ignoring a potentially vital component of our experiment. In this respect, there are two extensions of the Hertzian model, the Derjaguin-Muller-Toporov (DMT) [47, 48] and Johnson Kendall-Roberts (JKR) [25, 26] model. While the JKR model considers adhesion forces in the area of contact only, the DMT model is ruled out since it takes only long-range adhesion forces apart from the contact area into account. Hertzian models assume the probe as a perfect sphere with radius $r$, which equals the AFM tip radius, interacting with a perfect plane. The Tabor parameter [49] is a useful tool which allows selecting the correct model for our specific measurements. For a minimum tip-sample separation of around 0.5 nm, we estimate the Tabor parameter to be around $\mu = 5\text{-}20$, which again suggests that the JKR model is the preferred choice in our case.

The useful JKR relations for the contact radius $R$ and indentation depth $\delta$ respectively (as illustrated in **Fig. 2c**) are as follows:

$$R = \left(\frac{3r}{4E^*}\left(F_n + 6\pi\gamma r + \sqrt{12\pi\gamma r F_n + (6\pi\gamma r)^2}\right)\right)^{\frac{1}{3}} \tag{3}$$

$$\delta = \frac{R^2}{r} - \frac{2}{3}\sqrt{\frac{\pi\gamma R}{E^*}} \tag{4}$$

where *r* denotes the tip radius (in our case *r* = 60 nm [16]), ɣ the work of adhesion, $F_n$ the normal loading force, and $E^*$ the effective Young's modulus.

## 5. Topography and mechanical phase images of a PMMA sphere submerged in water

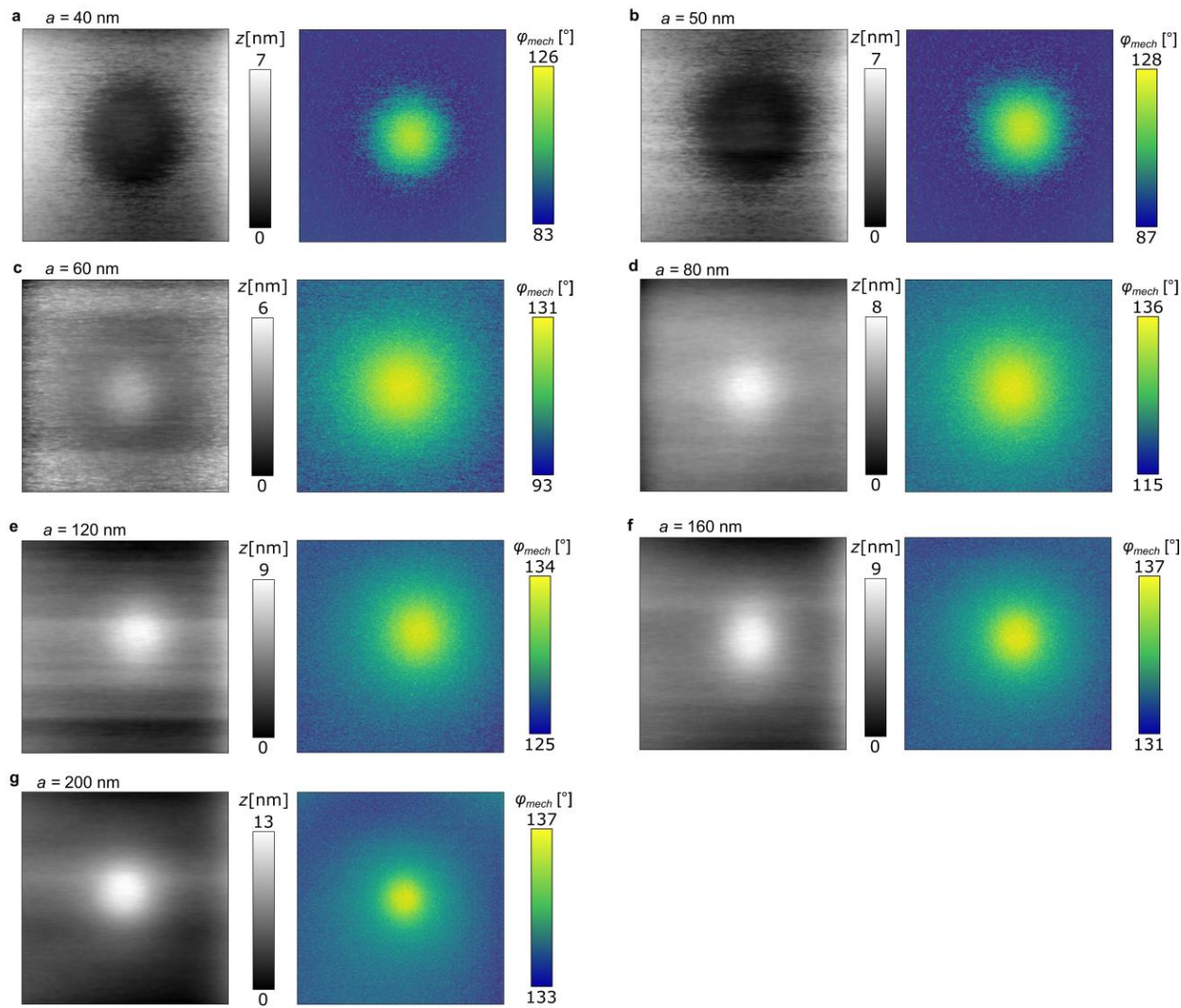

**Figure S3. Measured topographies and tapping phase images of PMMA sphere underneath SiN membrane for different tapping amplitudes. a** 40 nm. **b** 50 nm. **c** 60 nm. **d** 80 nm. **e** 120 nm. **f** 160 nm. **g** 200 nm.

## 6. Mechanical images of a PMMA sphere in a dried environment two days after sample preparation

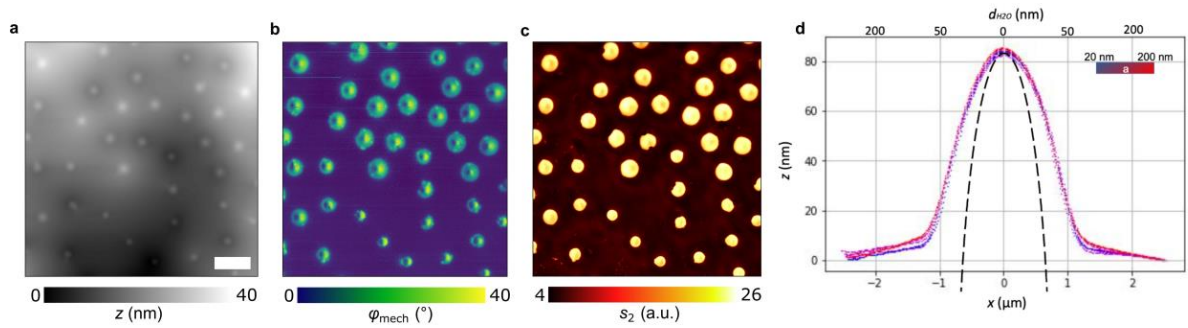

**Figure S4. s-SNOM mechanical images of a dried-out SiN membrane with PMMA spheres adhering underneath a** Topography, **b** tapping phase and **c** optical near-field amplitude $s_2$ taken on a 10-nm SiN membrane supported by a dried-out suspension of PMMA spheres supporting the membrane. The images were taken two days after sample preparation (same sample as in Fig. 3) with no water left (as has been verified through spectra not shown here). The membrane has experiences strong local deformations across the entire surface, which occurred during the drying process and were not present on the day of sample preparation. Scale bar 10 μm. **d** Topographical profiles taken along a single dried PMMA bead, showing no difference in deformation for varying tapping amplitudes and demonstrating a constant bulge of around 80 nm. We attribute this to the formation of a layer of small nanoparticles contained in the suspension that aggregate at the membrane surface during water evaporation and provide stability, preventing the membrane to be pushed downwards by the tip.

## 7. Extracted Profiles and Contrasts for different tapping amplitudes and amplitude setpoints

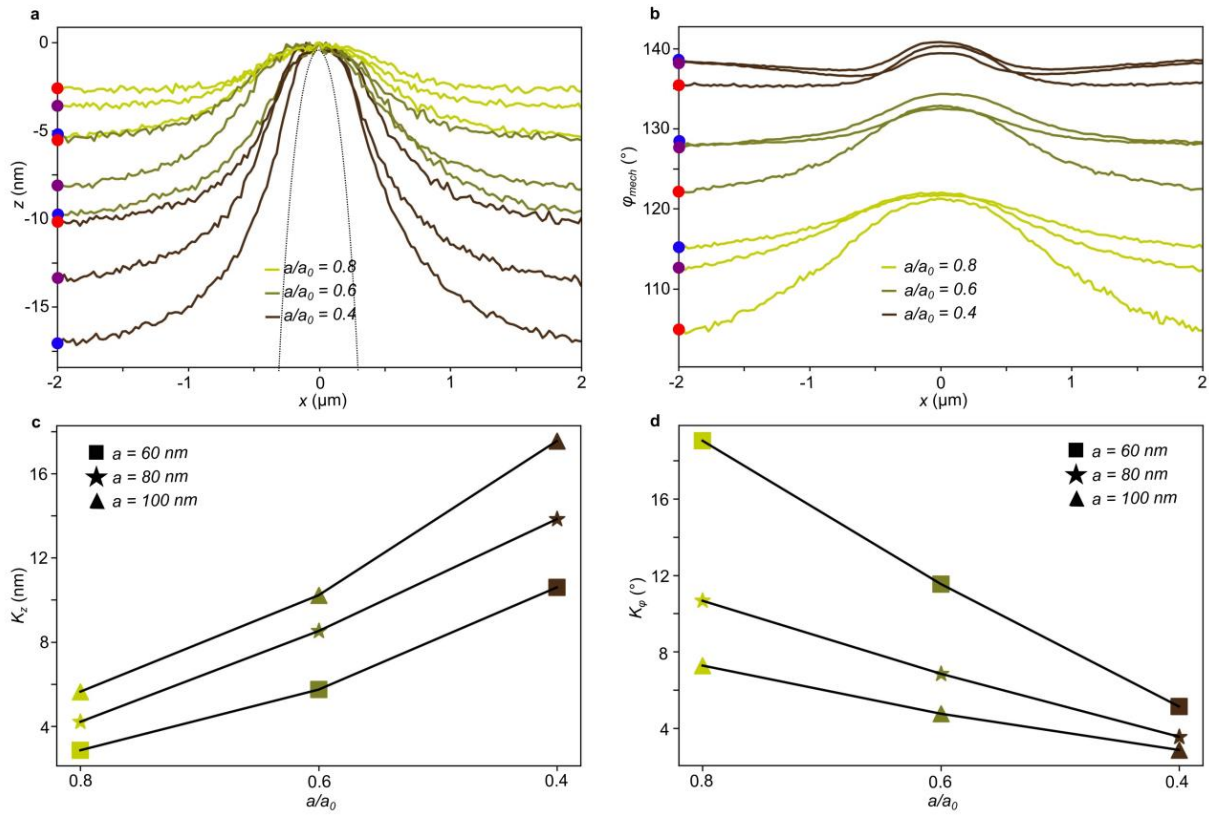

**Figure S5 Extracted Profiles and Contrasts for different tapping amplitudes and amplitude setpoints a, b** Extracted curves from Fig. 6 of topography and tapping phase respectively measured with different amplitude setpoints (yellow for $a/a_0 = 0.4$, dark yellow for $a/a_0 = 0.6$, brown for $a/a_0 = 0.8$), each measured with three different tapping amplitudes $a$ of 60 nm, 80 nm and 100 nm. Scale bar 1 µm. **c, d** Contrast between measured topography and tapping phase far away (2 µm) from the sphere and on top of the sphere as a function of amplitude setpoint for different tapping amplitudes $a$ (square shapes for $a$ = 60 nm, star shapes for $a$ = 80 nm, triangle shapes for $a$ = 100 nm). All data was acquired using a probe with radius $r$ = 60 nm.

## 8. Single-wavelength imaging of PMMA spheres in water

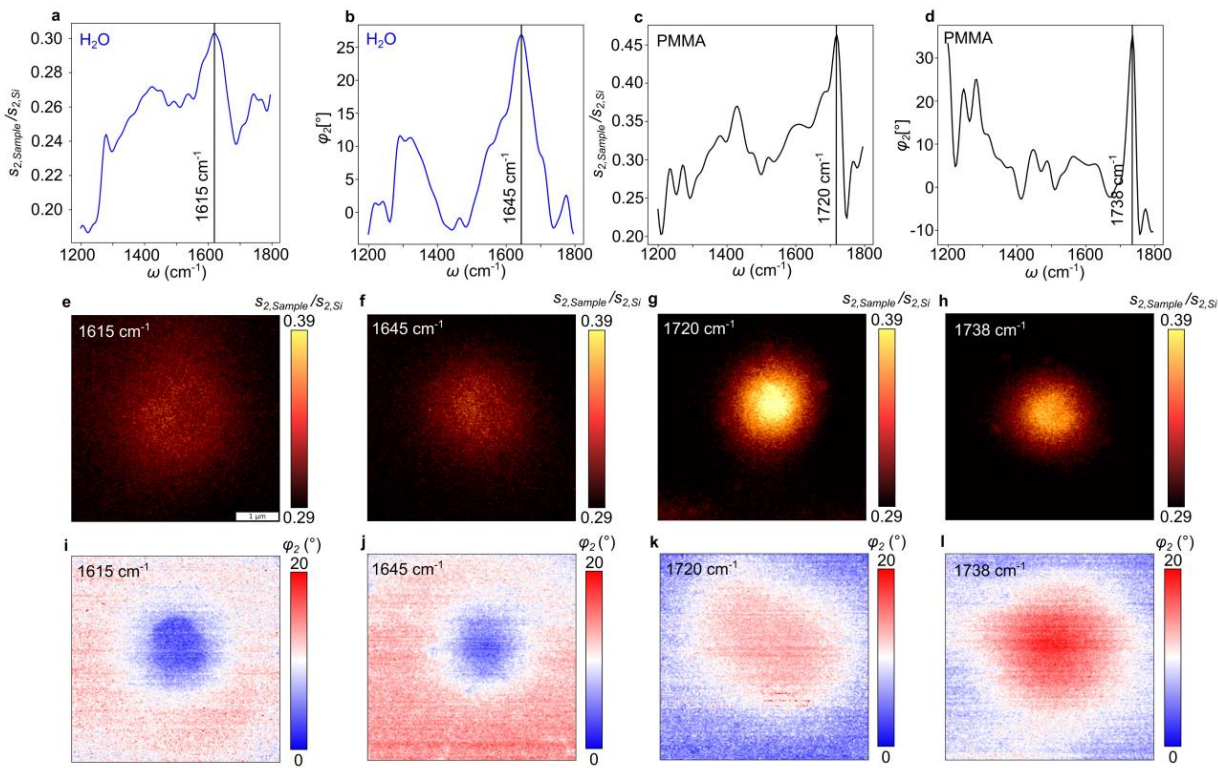

**Figure S6. Spectra and nano-imaging of PMMA beads in water a, b** Nano-FTIR second demodulated amplitude and phase spectra of water underneath a 10 nm SiN membrane. **c, d** Nano-FTIR second demodulated amplitude and phase spectra of a PMMA sphere underneath a 10 nm SiN membrane. Spectra were taken at the center of the sphere to ensure that no water layer is present. **e-l** Single-wavelength images of second demodulated optical amplitude and phase taken at the indicated wavelengths, illustrating contrasts that match the spectra shown in a-d.

## 9. Imaging of aggregated amyloid-beta protein strands in water

A frozen suspension containing amyloid-beta peptides was thawed at 0°C for 1/2 h, then 1:20 diluted with water before being pipetted it into the liquid cell (see **Experimental section**), then sealed after a deposition time of 45 min. s-SNOM images taken in 12 min sequence exhibit a rather compact 1x0.5 µm$^2$ protein aggregate that was stable in time for hours. Interesting is the high spatial resolution attained, of <30 nm as can be estimated from repeated detail in grainy structures around the object (pixel size about 15 nm). The body of the aggregated object exhibits a rather flat topography and likewise, quite homogeneous contrasts in both infrared and $\varphi_{mech}$ images, as compared to the surrounding.

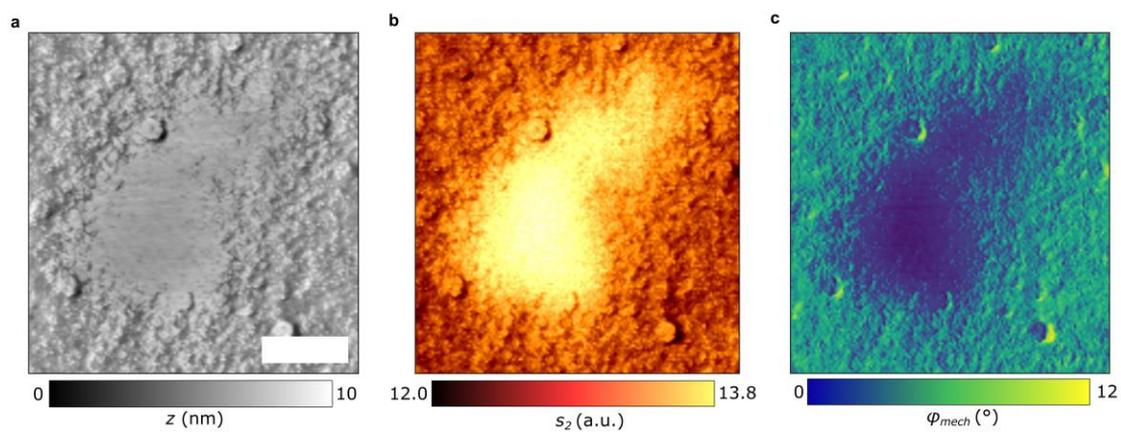

**Figure S7. s-SNOM images of amyloid-beta protein aggregates in water a** Topography, **b** $s_2$ infrared amplitude ("white light" spectral average), **c** tapping phase of aggregated amyloid-beta protein strands adhering in water to a 10-nm SiN membrane, tapping amplitude a = 80 nm. Scale bar 500 nm.